\RequirePackage{fix-cm}
\documentclass[twocolumn,epjc3]{svjour3}  
\smartqed  
\RequirePackage{graphicx}
\usepackage{graphicx}
\graphicspath{ {./figures/} }
\usepackage{siunitx}
\usepackage{listings}
\usepackage{color}
\usepackage{soul}
\usepackage[frozencache]{minted}
\usepackage{amsmath}
\usepackage{amssymb}
\usepackage{xcolor}
\usepackage{hyphenat}
\usepackage{xcolor}
\definecolor{darkred}{rgb}{0.5,0,0}
\definecolor{darkblue}{rgb}{0,0,0.5}
\definecolor{firebrick}{rgb}{0.75,0.125,0.125}
\definecolor{darkgreen}{rgb}{0,0.5,0}
\usepackage[colorlinks=true,linkcolor=firebrick,citecolor=darkgreen,urlcolor=darkblue]{hyperref} 

\newcommand*{\doi}[1]{\href{https://doi.org/\detokenize{#1}}{DOI: \detokenize{#1}}}
\journalname{Eur. Phys. J. C}
\begin{document}
\title{NuRadioReco: A reconstruction framework for radio neutrino detectors}

\author{Christian Glaser\thanksref{e1,uci}
        \and
        Anna Nelles\thanksref{e2,desy,fau}
        \and
    Ilse Plaisier\thanksref{fau,desy}
    \and
    Christoph Welling\thanksref{fau,desy}
    \and
Steven W.~Barwick\thanksref{uci} 
\and
Daniel Garc\'ia-Fern\'andez\thanksref{fau,desy}
\and
Geoffrey Gaswint\thanksref{uci}
\and
Robert Lahmann\thanksref{fau, uci}
\and
Christopher Persichilli\thanksref{uci}
}

\thankstext{e1}{e-mail: christian.glaser@uci.edu}
\thankstext{e2}{e-mail: anna.nelles@desy.de}


\institute{Department of Physics and Astronomy, University of California, Irvine, CA 92697, USA\label{uci}
           \and
          DESY, Platanenallee 6, 15738 Zeuthen, Germany \label{desy}
           \and
           Erlangen Centre for Astroparticle Physics, Friedrich-Alexander-Universit\"at Erlangen-N\"urnberg, 91058 Erlangen, Germany \label{fau}
}

\date{Received: date / Accepted: date}

\maketitle

\begin{abstract}
While the radio detection of cosmic rays has advanced to a standard method in astroparticle physics, the radio detection of neutrinos is just about to start its full bloom. The successes of pilot-arrays have to be accompanied by the development of modern and flexible software tools to ensure rapid progress in reconstruction algorithms and data processing. We present NuRadioReco as such a modern Python-based data analysis tool. It includes a suitable data-structure, a database-implementation of a time-dependent detector, modern browser-based data visualization tools, and fully separated analysis modules. We describe the framework and examples, as well as new reconstruction algorithms to obtain the full three-dimensional electric field from distributed antennas which is needed for high-precision energy reconstruction of particle showers.
\end{abstract}


\section{Introduction}
In this article, we present a novel modular framework for the detector simulation and data reconstruction of radio detectors for neutrinos and cosmic rays along with the corresponding algorithms. 
For neutrino detection, the radio technique allows to significantly extend the energy range of current experiments of a few times \SI{e15}{eV} \cite{IceCubePRL,IceCubePRL_E}, which is required to reach the next major milestone in astroparticle physics: the discovery of cosmogenic neutrinos \cite{1966PhRvL..16..748G,1966JETPL...4...78Z,1969PhLB...28..423B}. 
In high-energy cosmic-ray physics, the radio technique has already been established as a competitive detection method during recent years \cite{reviewHuege,reviewSchroder}.  In particular its excellent sensitivity to the cosmic-ray mass \cite{LofarNature} and energy \cite{AERAPRD,AERAPRL} make this technique very attractive. While rivalling in accuracy it is less sensitive to atmospheric conditions than optical methods and has a duty cycle of close to 100\% \cite{LOFARrefractiveindex,Gottowik_2018}. 

Many aspects of data processing, detector simulation and reconstruction are similar between radio detectors for cosmic rays and neutrinos.  Radio neutrino detectors such as ARIANNA are even cosmic-ray detectors themselves \cite{ARIANNACRs}. Hence, many analysis methods and strategies from the cosmic-ray community can be transferred to neutrino detectors allowing to benefit from the maturity of radio cosmic-ray observations. Consequently, it was the obvious choice to develop a framework suitable for both neutrinos and cosmic rays. 

This framework builds on extensive experience with both Monte-Carlo studies and data analysis of cosmic-ray as well as neutrino detectors \cite{LofarNature,AERAPRD,AERAPRL,ARIANNACRs,GlaserErad2016,LOFAREnergy}. It is also based on many years of experience with the software needs of large radio cosmic-ray experiments, in particular the experience with existing software projects such as Offline \cite{OfflineSoftware,RadioOffline}, the reconstruction framework of the Pierre Auger Observatory and its radio extension AERA, the LOFAR cosmic-ray software \cite{2013A&A...560A..98S}, the Physics eXtension Library (PXL) for high-energy phy\-sics and the web analysis framework VISPA \cite{VISPA}. NuRadioReco combines their strengths while addressing shortcomings of the existing projects for radio detection. 

NuRadioReco was developed in the context of the ARIANNA \cite{ARIANNA2015}, a pilot-array for the detection of high-energy neutrinos with energies above \SI{e16}{eV}. It consists of an array of autonomous stations located close to the surface on the Antarctic ice sheet. Each station has multiple spatially separated antennas with different orientations to reconstruct the incoming signal direction and polarization. Radio signals are produced via the Askaryan effect \cite{Askaryan} from particle cascades generated in the ice by interactions of these neutrinos. The Antarctic ice is transparent to MHz--GHz radio signals which allows for a cost-effective instrumentation of large volumes \cite{barwick_besson_gorham_saltzberg_2005}. Therefore, the radio technique is the method of choice with two fully-operational pilot detectors \cite{ARA,ARIANNA2015} and larger experimental efforts planned for the near future \cite{RNO,COSPAR}.

Established cosmic-ray detectors such as AERA \cite{AERAPRD}, LOFAR \cite{LOFAR} and Tunka-Rex \cite{TunkaRex} also consist of many autonomous detector stations. Here, each station has just one dual-polarized antenna and a coincident measurement of multiple stations is required for data analysis, while neutrino detectors typically only have radio signal data from one station, however with multiple antennas. The necessary flexibility to account for either is part of NuRadioReco. 

This article serves two purposes: First, to document NuRadioReco and second, to describe the algorithms required to reconstruct data from radio neutrino detectors, in particular the algorithms to recover a radio signal from multiple spatially displaced antennas with different polarization responses, a problem not yet addressed in literature. An accurate reconstruction of the electric field is the foundation for a high-precision measurement of the  energy contained in a neutrino or cosmic ray induced particle shower using its radio emission.

\begin{figure*}[t] 
\centering
\includegraphics[width=0.8\textwidth]{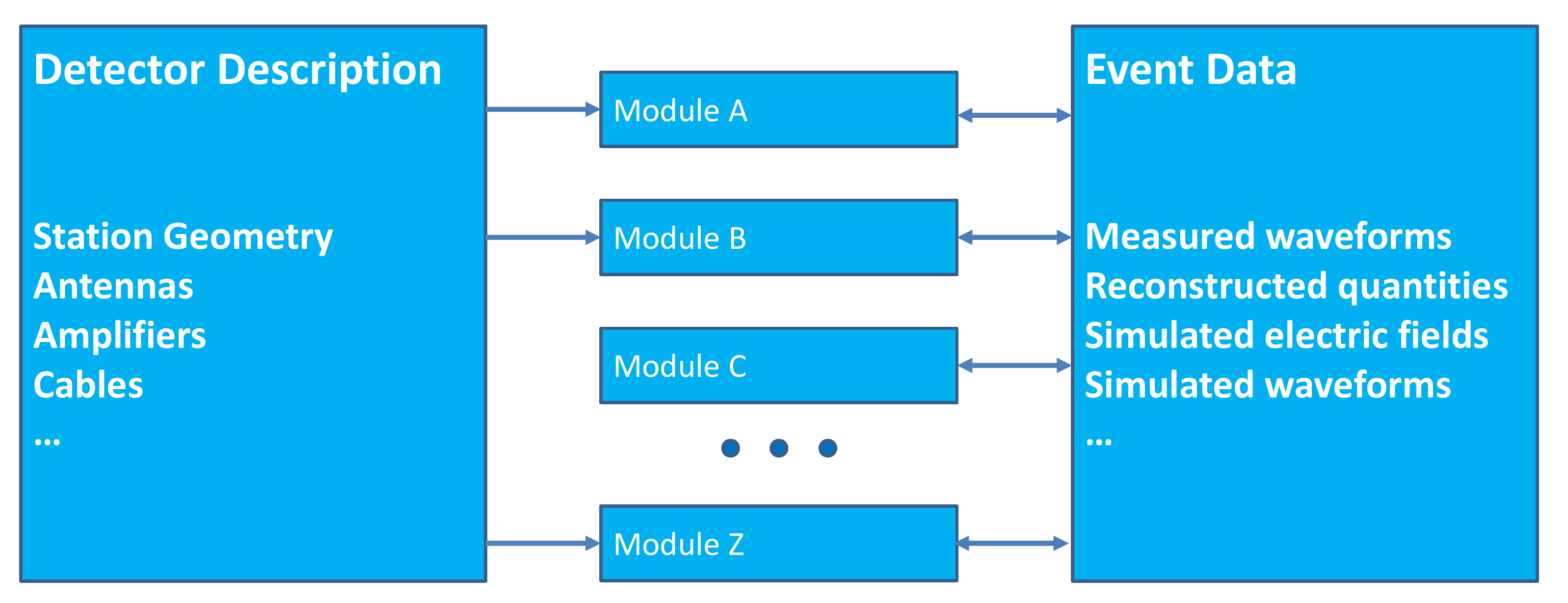}
\caption{The three principal parts of NuRadioReco. Detector description and event data are strictly separated. The data-analysis is carried out through fully separated processing modules. }
\label{fig:structure}
\end{figure*}

NuRadioReco is written in Python, open-source and publicly available on github \cite{NuRadioReco}. The design goals are to be easy-to-use with a user friendly interface and a maximum amount of flexibility. It follows a modular design with a strict differentiation between event data, detector description and processing modules depicted in Fig.~\ref{fig:structure}. The breakdown of the data processing into independent steps (the processing modules) fosters collaborative development, enforces a clear structure and allows for an easy modification of a processing pipeline. Each module is independent of each other, as modules only interact with the event data, and thus can be exchanged easily.  Through the consistent use of numpy \cite{numpy} for all operations on arrays, the code is sufficiently fast while offering all advantages of Python: Easy software installation as no compilation is required which is often cumbersome on different systems and platforms. The flexible Python steering allows arbitrary loops around modules, complex if/else branches, stop criteria and the possibility to call a module several times with different arguments. At the same time, more complex calculations can be included in optimized compiled languages such as C++, as long as Python wrappers are provided. 

In the following paragraphs, we briefly discuss the main advantages of NuRadioReco. 
The details are given in the individual sections of this paper. We also describe the properties of all default modules and provide an end-to-end example of a signal simulation, full detector simulation, and data reconstruction. The paper concludes by describing two new algorithms used in signal processing for single-station detectors.

\paragraph{I/O}
In NuRadioReco, the default input and output file format is the same. The i/o modules allow to save the current state of the event data to disk after each processing step, and to read it back in. Hence, a data processing pipeline can be split up into consecutive steps. It is also straight forward to implement modules to read instrument generated data into the event structure. 

\paragraph{Time-dependent detector description}
In NuRadioReco we use an SQL database designed to store a time-de\-pen\-dent detector description. While SQL is the method of choice to store the description of a large experiment, SQL has its limitations in usability and for queries from parallel processing on clusters. Therefore, we implemented a database export into a human readable JSON text file. This also allows for a simple setup of new detectors for simulation studies. 

\paragraph{Data visualization - event browser} 
Data is visualized using state-of-the art web technologies. The GUI is platform independent as the only requirement is a web browser. This design also allows for a remote deployment such that data can be inspected over the internet. This is particularly useful for outreach activities and easy collaborative sharing of data and results.

\paragraph{Default system of units} 
Keeping track of units is a must for physics analyses. NuRadioReco employs the same concept as \cite{OfflineSoftware}: Every time a variable is defined, it is multiplied by its unit, and every time a variable is plotted or printed out, it is divided by the unit of choice, such as  
\begin{minted}[
fontsize=\footnotesize
]{python}
from NuRadioReco.utilities import units
time = 132. * units.ms # define 132 milli seconds
d = 5. * units.mm # define 5 mm
v = d/time  # calculate speed
print("the speed is {:.2f} km/h".format(v/units.km* 
        units.hour))
# the speed is 0.14 km/h
\end{minted}
In this way, all internal computations can be done without the need for the user to worry about the correct units. We have chosen to not import an existing unit system (such as astropy or pypi), as we need access to units from all modules, including those not written in Python.

\paragraph{Link to simulations}
Simulations of the radio emission following a neutrino interaction are currently performed with limited flexibility in detector design and using simple signal parameterizations only (e.~g.\ \cite{ShelfMC,ARASim}). In parallel to NuRadioReco, NuRadioMC is being developed as community-driven simulation code that addresses the short-comings of previous codes \cite{NuRadioMC}. As it shares certain characteristics of NuRadioReco, such as the data-format, there will be a seamless integration of signal simulation, detector simulation and reconstruction for future experiments. 

\section{Data structure}
All measured, simulated and reconstructed quantities are saved in a hierarchical class structure that also supports the simple storage of analysis quantities and is designed having the (for radio experiments) natural representation in time- and frequency-domain in mind. 

\subsection{Event structure}
The class structure fits the requirements of multi- and single-station detectors and both cosmic-ray and neutrino reconstruction. Askaryan neutrino detectors, such as ARIANNA \cite{ARIANNA2015} and ARA \cite{ARA}, consist of independent detector stations, i.e., the design foresees that measurement, identification and reconstruction of a neutrino properties are done using data from a single station. In contrast, typical cosmic ray detectors, such as AERA \cite{AERAPRD} or LOFAR \cite{LOFAR}, consist of many stations that collectively measure the cosmic-ray signal. Although multi-station coincidences are not typical for Askaryan neutrino detectors, very high energy events or `double-bang' tau events might be observed in multiple stations. Hence, the data structure is flexible enough to accommodate both cases. However, the focus of this paper lies on single station events. 

The treatment of radio signals from air showers (cosmic rays) and in-ice showers (neutrinos) is slightly different. The air-shower signal is measured at large geometrical distances to the shower maximum and thus extends over a large area. It can safely be assumed that the signal does not change over the small lateral extent of a compact station of a few meters\footnote{This is technically only true at low frequencies, but an elaborate discussion of the emission at high frequencies would go beyond the scope of this paper.}. Hence, all antennas of one station observe the same signal. This is not the case for the Askaryan signal of in-ice showers. Here, the showers are observed at closer distances in an inhomogeneous medium. Thus, the signal can be significantly different, especially for antennas displaced in depth, because of its strong dependence on the viewing angle of the Askaryan signal. Furthermore, each antenna may detect two pulses from the same in-ice shower from different directions and propagation paths through the ice \cite{KelleyARENA2018}. This is because the upper ice layer is a non-uniform medium where the signal trajectory is bent, leading to two distinct solutions, either a direct path, a refracted path or path where the signal gets reflected off the ice-air interface at the surface. Also more exotic emission models might lead to even more pulses per antenna with different incoming directions. Consequently, there is the need to store an arbitrary number of signals that arrive at the same channel at different times and from different directions. 

\begin{figure*}
\centering
\includegraphics[width=0.65\textwidth]{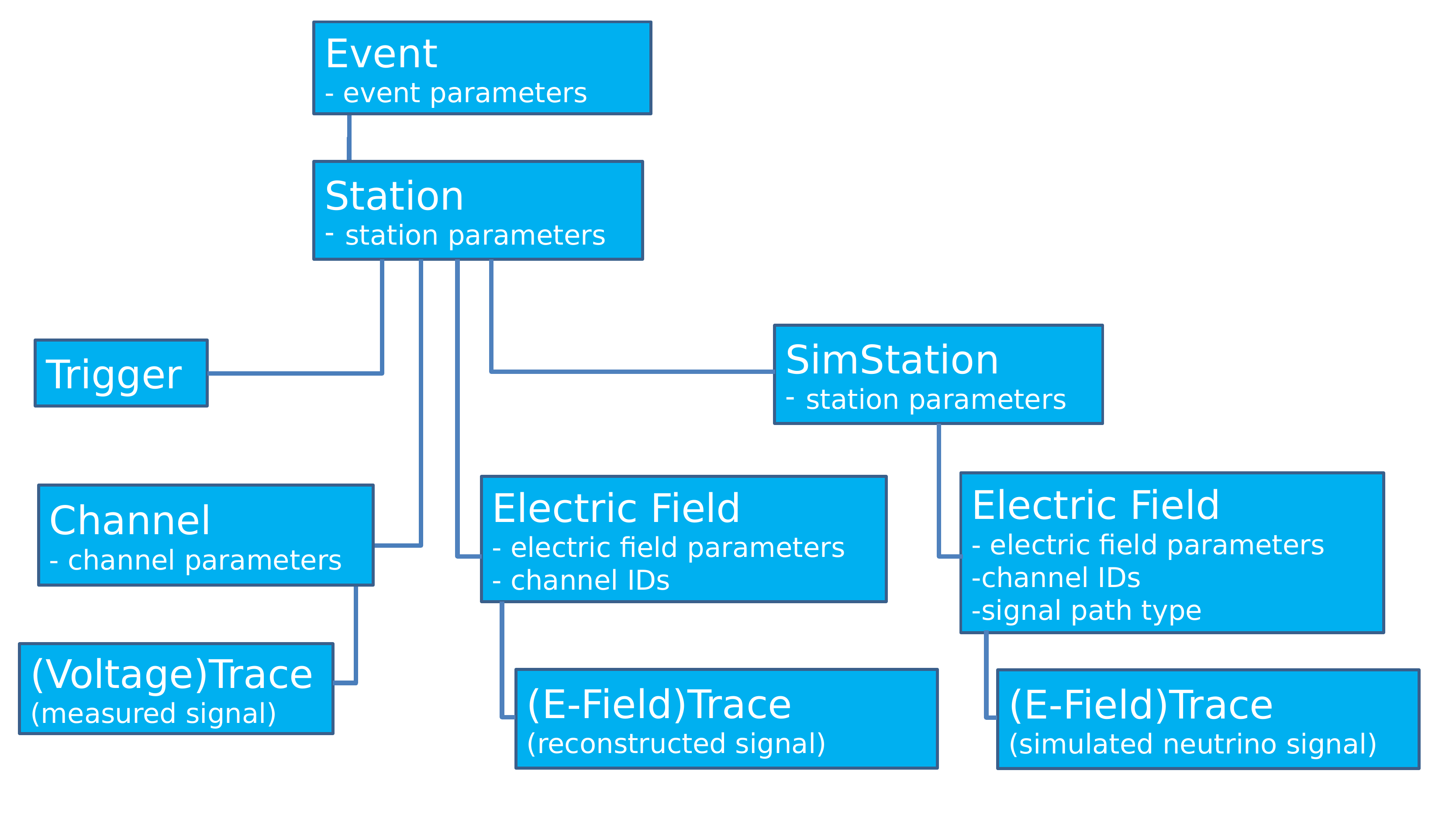}
\caption{Event structure as used in NuRadioReco. NuRadioReco uses the event as basic hierarchical structure. Every event contains stations and can be broken down to the waveforms in the voltage traces. }
\label{fig:eventstructure}
\end{figure*}

All these requirements can be mapped into the event structure depicted in Fig.~\ref{fig:eventstructure}, also showing the definition of hierarchical levels. First, the \emph{event} level that includes all simulated, measured and reconstructed data of all stations that have detected a signal. Second, the \emph{station} level that includes all antennas of a single radio detector station, and third, the \emph{channel} level, one for each antenna, storing the measured signal. All simulated quantities are stored in the \emph{SimStation} class. 

Furthermore, we differentiate between voltage traces $V_i(t)$, i.e., the signal as a function of time measured in an antenna $i$, and electric-field traces $\vec{E}(t)$ that refer to the three-dimensional electromagnetic pulse before being measured by the antenna. Electric fields are stored in a dedicated \emph{electric field} class. Apart from storing the time series, it also stores the incoming signal direction and the information for which channel(s) it is valid. This allows to cover the cosmic-ray case, i.e., a single electric field is valid for all channels, as well as the neutrino case where it might be necessary to store an electric field for every channel separately. For simulated neutrino events, we can have multiple electric fields per antenna from different signal paths that are treated by adding a second electric field for the same channel. 

\subsection{Parameter storage}
The data structure offers a flexible mechanism to save parameters on event, station and channel level, e.g., the reconstructed air-shower direction on station level and the signal-to-noise ratio on channel level. All parameters are defined in an enumerated type \emph{enum} and can be accessed via a generic \emph{setter} and \emph{getter} function. We also allow to save both an uncertainty for each parameter and the covariances between any pair of parameters.

To add a new parameter, the parameter simply needs to be added to the \emph{enum} table. Then, this parameter can be accessed from each module and is automatically included in the input/output file. This is an advantage in time-efficiency compared to the standard way of adding a new member variable for each parameter, because in the latter case additional \emph{getter} and \emph{setter} methods need to be implemented and the variable needs to be manually included into the i/o data stream. 

Explicitly defining all parameters in an \emph{enum} ensures that all parameters are well defined and that users know which parameters exist. Therefore, we chose not to use a Python dictionary to store parameters but implemented a dictionary-like usage:

\begin{minted}[
fontsize=\footnotesize
]{python}
from NuRadioReco.framework.parameters 
    import stationParameters as stnp
from NuRadioReco.utilities import units

# set parameters via generic setter function
station.set_parameter(stnp.nu_energy, 7e8 * units.GeV)
# or via dictionary like interface
station[stnp.nu_energy] = 7e8 * units.GeV
# set uncertainty of neutrino energy
station.set_parameter_error(stnp.nu_energy, 
    1e6 * units.GeV)
# access of parameters 
nu_energy = station.get_parameter(stnp.nu_energy)
# or
nu_energy = station[stnp.nu_energy]
\end{minted}

\subsection{Time and frequency domain}
The voltage and electric-field traces can be represented in the time or frequency domain. The two representations can be used interchangeably as depending on the processing step one representation may be more convenient to work with. For example, a bandpass filter is implemented easiest in the frequency domain, whereas a pulse finding algorithm is naturally implemented in the time domain. Therefore, the event structure offers the functions
\begin{minted}[
fontsize=\footnotesize
]{python}
# access trace in time domain
time_trace = channel.get_trace()
times = channel.get_times()
# or access trace in frequency domain
frequency_spectrum = channel.get_frequency_spectrum()
frequencies = channel.get_frequencies()
\end{minted}
to transparently obtain the time or frequency domain representation depending on the needs of a processing module. Internally, it is kept track of which representation was last modified and a Fourier transform is performed if necessary. Similarly, we provide functions to define a new trace either in time or frequency domain. Technically this functionality is implemented once in a generic \emph{base trace} class from which the \emph{channel} and \emph{electric field} classes inherit. 

This approach avoids typical errors using Fourier transforms and their normalization. We chose to normalize the transforms as such that Parseval's theorem is observed and the physical quantity of signal power is conserved.

\section{Input/Output}
NuRadioReco provides several input modules for different sources (CoREAS simulations \cite{Coreas}, ARIANNA raw data format, etc.) but also has its own file format, ending by default in \emph{*.nur}. 

\subsection{Philosophy of \emph{.nur} files}
The main advantage of NuRadioReco's own \emph{.nur} file format is that it was designed to save/read the current state of the event data to/from disk after every modular processing step. Therefore a reconstruction can be split up into multiple steps without complications. For example, computationally expensive low level processing only needs to be done once and secondary reconstruction can be started from the pre-processed files without having to save data in an intermediate file format. 

A practical application for the ARIANNA experiment is the following. Most triggers are caused by thermal noise fluctuation with typical rates of a few events per minute per detector station. The rate of cosmic rays is only one or two per day. Hence, in a first processing step, cosmic-ray candidate events are identified and only those events are saved to disk. This largely reduces the data volume to an easily manageable file size and serves as the starting point of high-level analyses. 

\subsection{Technical implementation}
The NuRadioReco data format is implemented through a serialization and deserialization function in each event data class, a concept adapted from PXL \cite{VISPA}. In other words, the data structure knows how to (de)serialize itself. The (de)serialization is performed recursively per event, i.e., calling the serialization function of the event class will call the serialization function of all stations that are part of this event, which will call the serialization function of its channels and so on. Another advantage is that new properties can be added to the event data structure without the need to also modify the i/o-modules. A new property only needs to be added to the (de)serialization function. 
This implementation also allows for backward compatible additions to the file format. If a certain property is not present in an older file version, it can be initialized with an appropriate default value during the deserialization. 

Internally, it is made use of the \emph{pickle} module to create a binary representation of most data members, but the data file itself is a custom binary format and the use of \emph{pickle} could be replaced in the future. The data format is compatible across different computing systems and Python versions.  The only requirement to read the data is a Python installation and the NuRadioReco modules but no installation is required. We note that we also considered other file format options, e.g., to build our data format on top of HDF5 but didn't find it suitable for our case (see e.g. the discussion in \cite{HDF5discussion} and \cite{HDF5discussion2}).

In addition to the full storage of the event structure, the high-level parameters on station level are saved in an additional event header. This enables quick parsing of data files, and access and plotting of high-level quantities. During the initialization of the i/o class, the headers of all events are parsed and the high-level parameters are stored in numpy arrays. This allows for a quick inspection and plotting of analysis results. With just a couple of lines of code one can plot the maximum pulse amplitude as a function of time, or a histogram of the reconstructed signal directions as:

\begin{minted}[
%frame=lines,
%framesep=2mm,
%baselinestretch=1.2,
fontsize=\footnotesize,
%linenos
]{python}
import NuRadioReco.modules.io.NuRadioRecoio 
    as NuRadioRecoio
from NuRadioReco.utilities import units
from NuRadioReco.framework.parameters 
    import stationParameters as stnp
import matplotlib.pyplot as plt

nurio = NuRadioRecoio.NuRadioRecoio("my_file.nur")
header = nurio.get_header()
station_id = 51
station_header = header[station_id]

# get numpy arrays of reconstructed direction
zeniths_rec = station_header[stnp.zenith]
azimuths_rec = station_header[stnp.azimuth]

# plot zenith vs azimuth in degrees
plt.plot(zeniths_rec/units.deg, 
    azimuths_rec/units.deg) 
plt.show()
\end{minted}

Another feature of the NuRadioReco i/o class is that the amount of data written to disk can be controlled on an event-by-event basis. The vast majority of the disc space is typically occupied by signal data as voltage or electric field trace. NuRadioReco offers three output modes: 
\begin{itemize}
\item `full' (default): the full event content is written to disk
\item `mini': only electric-field traces are written to disc, but no channel traces
\item `micro': no traces are written to disc
\end{itemize}
that are specified in the \emph{eventWriter}'s run method, e,g., \mintinline{python}{eventWriter.run(evt, 'micro')}.

Another feature is that output files can automatically be split up into several files by specifying a maximum file size. Correspondingly, the \emph{eventReader} has the functionality to read in a list of files transparent to the user. 

\section{Time dependent detector description}
\label{sec:detector}

Any larger experimental effort requires a complete detector description that provides all information relevant for data analysis in a machine readable form. This includes the position and orientation of each antenna of each station, the details of the analog signal chain of each channel such as cable lengths, amplifier responses and ADC (analog-to-digital converter) details, and so on. Furthermore, a detector layout might change over time, detector stations might be reconfigured or certain components might be replaced. Hence, the detector description needs to be time-dependent. The requirement is that the user can request the exact configuration of a station at any time. 

\subsection{Database structure}
\begin{figure*}
\centering
\includegraphics[width=0.9\textwidth]{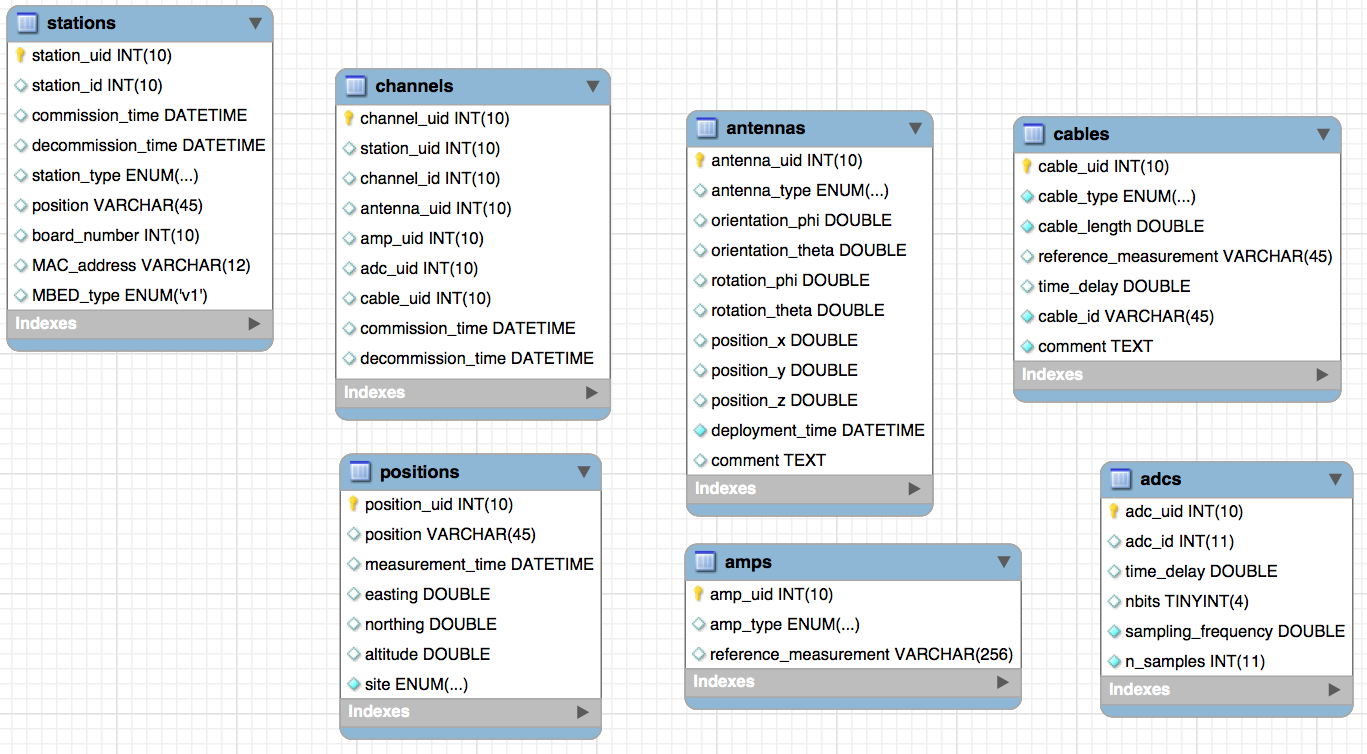}
\caption{MySQL database layout for the time-dependent detector description as used in NuRadioReco.}
\label{fig:mysql}
\end{figure*}

The method of choice is to store the detector description in a database. We use MySQL and present the database structure in Fig.~\ref{fig:mysql}. We have followed standard database design rules, in particular that no information is ever duplicated. Similar to the event data structure, the database has a hierarchical table structure. Different tables are related to each other by their unique ids. For example, to add a channel to a station, a new row needs to be inserted into the \emph{channels} table with the unique id of the respective station. Each \emph{channel} entry contains a reference to an antenna, cable, amplifier (amp) and ADC, which need to be defined in the corresponding tables. 

This design has a number of practical advantages: Each channel can have the same reference amplifier without re-specifying what \emph{reference amplifier} means. If the reference measurement changes, it needs to be changed only at one place. If a more detailed description is demanded for analysis, individual measurements of the amplifier response can be added to the \emph{amps} table and referenced from the respective channels.

The time dependent nature is implemented at two places. The \emph{stations} and the \emph{channels} table contain a commission and decommission time. A typical use case is that one or multiple channels of a station are reconfigured with a new antenna. To put this into the database, first the new antenna properties need to be added (as a new row) into the \emph{antennas} table. Then, the decommission time of the current channel is set to the time of the hardware change and a new channel is inserted into the \emph{channels} table with the same properties as the previous channel but with the antenna id pointing to the new antenna and with the proper commission and decommission time. 

An alternative way to implement the time dependence would have been to give each \emph{antennas}, \emph{cables}, \emph{amps} and \emph{adcs} entry a time dependence, and to remove the time dependence of the channel. And instead of referencing the channel components from the \emph{channels} entry, the \emph{antennas}, \emph{cables}, \emph{amps} and \emph{adcs} entries would reference back to the channel, similar to the \emph{stations} - \emph{channels} relationship. Such a structure would have the advantage that, for the above mentioned antenna replacement, only the antenna table needed to be altered. However, it comes with the large disadvantage that no default detector components can be specified. Suppose most ADCs are so similar that we can use the same ADC reference description for most channels. In the latter structure, a reference ADC entry needs to be added for every channel which is a huge duplication of information. Consequently, we have chosen the former design. 

The tables can be combined via 'JOIN' statements. The following code, for example, retrieves the position of the antenna of channel 2 of station 10 on November 5th 2018 at noon:
\begin{samepage} 
\begin{minted}[
%frame=lines,
%framesep=2mm,
%baselinestretch=1.2,
fontsize=\footnotesize,
%linenos
]{mysql}
SELECT position_x, position_y, position_z 
    FROM stations AS st
    JOIN channels AS ch USING(station_uid)
    JOIN antennas USING(antenna_uid)
WHERE CAST('2018-10-05 12:00' AS DATETIME) 
    between ch.commission_time and ch.decommission_time
    AND CAST('2018-10-05 12:00' AS DATETIME) 
    between st.commission_time and st.decommission_time
    AND st.station_id = 10 AND ch.channel_id = 2;
\end{minted}
\end{samepage}

\subsection{User friendly implementation}
Although a central database is the method of choice to keep track of the time-dependent detector description, it comes with several disadvantages for the user: 
\begin{itemize}
\item A (internet) connection to the MySQL server is required when running the software.
\item Queries to a remote MySQL server are relatively slow.
\item In MySQL the number of simultaneous connections to database is limited, which precludes parallel processing on computing cluster.
\item It is difficult to make local changes to the detector description for testing purposes.
\end{itemize}
Therefore, we have implemented additional options. Either the database is buffered at the beginning of the processing, i.e., we connect only once to the database, or the database is exported into a simple JSON text file 

The total amount of data required for the detector description is small, e.g., the complete detector description of the current ARIANNA detector is less than \SI{100}{kB}. Hence, we can buffer the complete database and store all information in memory. Internally, we use the \emph{TinyDB} Python package \cite{tinydb} to buffer the database. \emph{TinyDB} provides a convenient interface to the data and supports 'WHERE' statements to access the information of a specific station and channel at a specific time. However, \emph{TinyDB} does not support relationships between tables which we need to properly setup the detector description. Hence, we linearize the database and combine all channel related tables into one single table to store it in \emph{TinyDB}, which results in a duplication of data. At this point though, this is not a cause for concern as the master is always the MySQL database. 

\emph{TinyDB} also allows us to save the database in a simple JSON text file. This is the method of choice for most users and the default in NuRadioReco. A simple detector description is shipped as part of the software so that everything works out-of-the box. In this way, all the advantages of a MySQL database of storing a complex detector description are combined with the user-friendly usage of human readable text files. 

When running on a large cluster, the number of MySQL connections typically limits the number of parallel compute nodes. Through the usage of JSON files, this can be avoided by copying the relevant JSON files to individual nodes. Due to the limited size, i/o will also not be an issue.

We note that queries through \emph{TinyDB} are relatively slow even if all data is present in memory. Therefore, we have added an additional layer of buffering such that database access is a negligible fraction of the total processing time. 

\subsection{Usage for simulation studies}
NuRadioReco is not only used for data reconstruction of existing experiments but also used to simulate future experiments. As setting up a MySQL database has to be considered too much overhead for most simulation studies, the JSON text file representation offers a convenient method to quickly define arbitrary station configurations. A few examples of detector descriptions are included in NuRadioReco \cite{NuRadioReco} that can be adopted to the users needs. It should be noted that it is sufficient to specify only the relevant fields, e.g., if a simulation study does not simulate the ADC digitization, the ADC related tables can be left empty or can be removed completely from the JSON file. Custom detector descriptions are specified during the initialization of the detector:
\begin{minted}[
%frame=lines,
%framesep=2mm,
%baselinestretch=1.2,
fontsize=\footnotesize,
%linenos
]{python}
import NuRadioReco.detector.detector as detector
det = detector.Detector(
    json_filename="/path/to/my_detector.json")
\end{minted}
\subsection{Handling of antenna sensitivities}
NuRadioReco provides a convenient interface to antenna models and provides a library of commonly used antennas for neutrino and cosmic-ray detection. Most antenna models available in NuRadioReco have been simulated with WIPL-D \cite{Kolundzija2011}, but also antenna simulations using XFDTD \cite{XFDTD} or NEC-2 \cite{Nec2} are available. This is handled technically by pre-processing the raw antenna simulation output to the same data structure, stored in a pickle file. This also significantly reduces the file size from raw simulation output that can easily exceed \SI{1}{GB} for a fine sampling in frequency and incoming signal direction. The antenna response files are provided on a central server and are downloaded automatically when needed. The conversion scripts for WIPL-D and XFDTD are provided.

The antenna response is quantified as the vector effective length which is a complex quantity that depends on frequency and incoming signal direction and can be thought of as proportionality constant between the incident electric field and the voltage output of the antenna (cf. Eq.~\eqref{eq:H_full} for more details). However, this quantity is typically not a direct output of antenna simulation software, with the output differing from software to software. In \ref{sec:Antenna_effective} we detail how the output of the different simulations are converted into the relevant vector effective length.

NuRadioReco's antenna model class then provides a user-friendly interface. The requested antenna model is buffered in memory, it is interpolated to the required frequencies and angles, and all coordinate rotations are handled internally to match the orientation of the antenna in the detector description. 

\subsection{Coordinate system}
We differentiate between the relative coordinates of the components of a detector station and its absolute position. The positions of the components (e.g. the antennas) are expressed in a local Cartesian coordinate system with the coordinate origin in the horizontal center of the station and with $z = 0$ at the ice surface. The positive x-axis is oriented into the Easting direction. 

For stations at the South Pole we use a special coordinate system that moves with the ice (the ice drifts by approximately \SI{20}{m} each year) such that the station coordinates remain constant with time with respect to each other. For other locations we use the UTM coordinate system. Both absolute coordinate system are local Cartesian projections onto a 2D surface. Hence, standard euclidean geometry can be used to calculate e.g. distances between stations. Given the typical distance between stations of \SI{1}{km}, the Earth curvature can be neglected. However, it is foreseeable to introduce more a more refined treatment of coordinates in the future. 

\section{Data analysis and processing modules}
In this section, we describe the setup of the data analysis modules and briefly discuss the standard processing steps to extract the physics properties from radio data. This is illustrated by a full example of reading in a cosmic ray simulation, a detector simulation and the reconstruction. New techniques for data analysis beyond what is currently used in radio detection of cosmic rays and neutrinos are discussed in Sec.~\ref{sec:Reco_algo}.

\subsection{Format of data analysis modules}
A detector simulation and event reconstruction using NuRadioReco is split into several modules that are executed in sequence, with each module fulfilling one specific purpose. In principle, modules can be arranged in any order, including loop or if/else branching, though some may require a certain module to be executed beforehand; for example, using a module to apply a filter to an electric-field trace is only sensible if another module has reconstructed the field. 
\label{sec:example_rec}
\begin{figure*}
    \centering
    \includegraphics[width=0.8\textwidth]{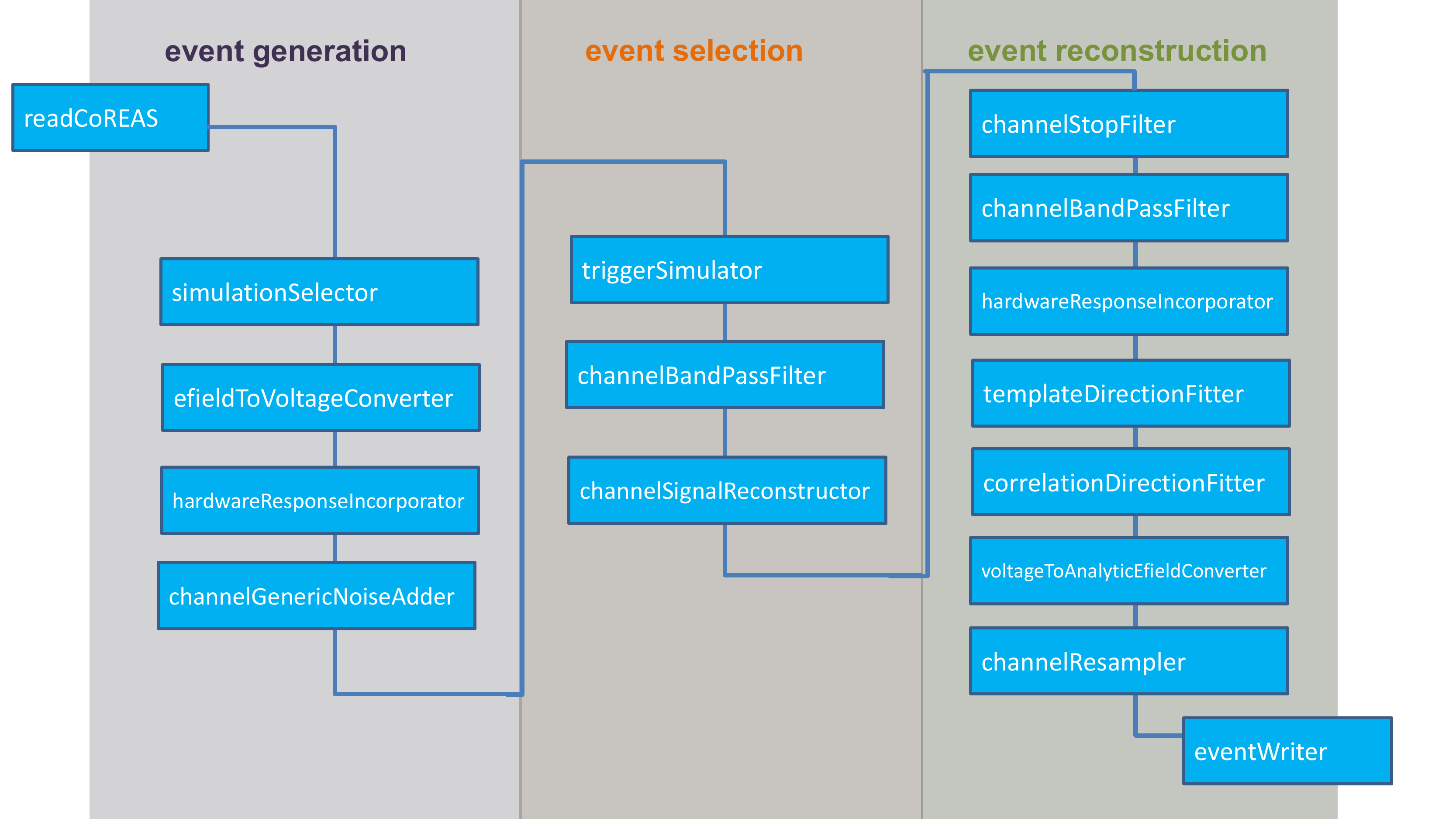}
    \caption{Schematic overview for a full reconstruction cycle for an event.}
    \label{fig:example_schema}
\end{figure*}

Each module consists of four components:
\begin{itemize}
    \item A constructor to create the module instance. This is called before the reconstruction loop over all events.
    \item A \emph{begin} function to set parameters that remain constant for each event, such as a list of input files for the event reader module.
    \item A \emph{run} function that is executed for each event and in which the module fulfills the task it was designed for.
    \item An \emph{end} function that may be executed after the last event was processed.
\end{itemize}

\subsection{Full simulation and reconstruction cycle}

For hands-on understanding of NuRadioReco, this section describes a full reconstruction cycle for a signal from an air shower simulated with CoREAS. A sche\-ma\-tic overview of all the modules used is shown in Fig.~\ref{fig:example_schema}. The full Python code, as well as a more detailed description are available at \cite{NuRadioReco_example}.

\subsubsection{Event generation}
\begin{figure*}
\centering
\includegraphics[width=0.8\textwidth]{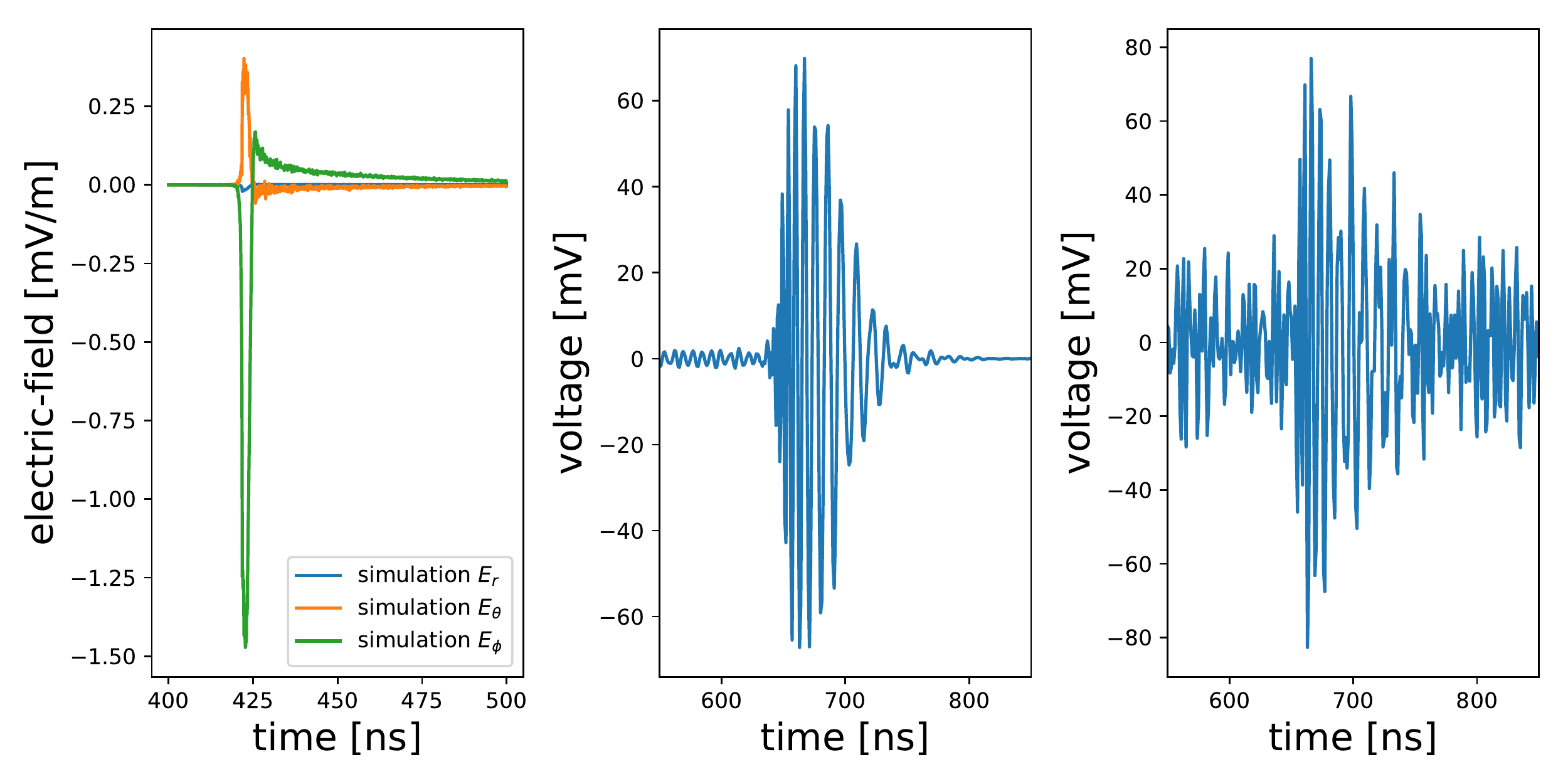}
\caption{The simulated electric-field traces (left), voltage trace in a channel resulting from full detector simulation (middle) and with noise added (right).}
\label{fig:example_traces}
\end{figure*}

The event generation starts by reading the electric-field traces that were simulated using CoREAS \cite{Coreas}. All events are selected that have signal in the frequency band of 100-500 MHz\footnote{CoREAS simulations show a mixture of numerical noise and incoherent signal at high frequencies that can mimic signal in the time-domain \cite{LOFARLDF}. Events with coherent signal in the relevant band are selected.}. In order to improve the timing accuracy, the simulated electric-field is up-sampled before the voltage traces per channel are calculated. This up-sampling is also needed to shift the signals in time according to the geometric time delays to the antennas with high precision. The ARIANNA hardware response and noise with an RMS of \SI{20}{\mV} is added to fully simulate data-like events. An example of the simulated electric-field traces, the calculated voltage trace for a channel, and the voltage trace with added noise are shown in Fig.~\ref{fig:example_traces}. 

\subsubsection{Event selection}
For the event selection, first a trigger is simulated, i.e., if a certain event would be recorded by the detector. Only the events marked as triggered are selected. In order for an event to trigger the amplitude must exceed \SI{100}{\mV} in at least two of the channels in this simple example. The voltage traces are filtered again in the frequency range of 80-500 MHz to reduce contamination by noise outside of the pass-band. 

\subsubsection{Event reconstruction}
The direction of the incoming cosmic ray is fitted using the direction-fitting modules. This reconstructed arrival direction is used to convert the voltage traces per channel to the fitted reconstructed electric-field. The electric-field reconstruction is done using the \emph{forward folding technique}, which uses an analytic description of the electric-field pulse and is discussed in detail in Sec.~\ref{sec:Reco_algo}. The simulated voltage traces and the reconstructed electric-field traces are down-sampled to the original detector bandwidth before storing to disk in order to reduce the file size. 

\subsection{Overview of relevant default modules}
A number of modules are considered default in NuRadioReco and many are used in the example. They are meant as illustration in the same way as starting point for more complex analyses. 

\paragraph{Event reading/writing}
The two modules \emph{eventReader} and \emph{eventWriter} allow to read and write from/into the NuRadioReco file format. 

\paragraph{Reading CoREAS files}
CoREAS is the state-of-the art simulation code for radio emission from air showers \cite{Coreas} and NuRadioReco provides a direct interface to the HDF5 output of CoREAS. Two different modules are provided serving different user requirements. The \emph{readCoREASStation} module creates a new event with a single station for each simulated observer in the CoREAS file, which is suitable for a detector like ARIANNA where air showers are measured independently by each station. 

The other module called \emph{readCoREAS} is optimized to read in 'star pattern' simulations for single station detectors such as ARIANNA. The spatial distribution of the radio emission can be sampled efficiently in a special coordinate system in the shower plane where one axis is oriented perpendicular to the air-shower axis and the geomagnetic field (see e.g. \cite{LOFARLDF,Alvarez-Muniz:2014wna,GlaserErad2016,Glaser_2019} for more details). Hence, instead of running a new time-consuming CoREAS simulation for different locations of the air shower on the ground, called \emph{shower core} in the following, we can just use one star pattern simulation and pick the closest station. This allows to reuse the same CoREAS simulation many times. 

The \emph{readCoREAS} module will generate a definable number of core positions randomly distributed on the ground. The module then determines the closest simulated station (measured in the shower plane) and creates and returns a corresponding event object. In this way, a realistic distribution of cosmic-ray events is obtained. 

\paragraph{Re-sampling}
Recorded traces are usually up-sampled at the beginning of the reconstruction process in order to improve accuracy and down-sampled after the analysis to reduce file size. This is done by using the modules \emph{channelResampler} to re-sample voltage traces and \emph{electricFieldResampler} to re-sample electric-field traces.

\paragraph{Band-pass filtering}
There are two modules available to filter signals: The \emph{channelBandPassFilter} for voltage traces and the \emph{electricFieldBandPassFilter} for electric field traces. Both support several different filter types, like a simple rectangular filter that cuts off all frequencies outside of its pass-band or a Butterworth filter modelling a hardware filter. Both can be applied to any frequency band.

\paragraph{Converting electric fields to voltages}
In order to perform simulation studies, it is necessary to calculate the waveform that an electric field produces in each channel, which is done with the \emph{efieldToVoltageConverter}. Since the radio pulse from a particle shower in the ice may reach the antenna via several different ways, each channel may have multiple electric fields associated with it, each resulting from a different ray path. Therefore, the first step is to calculate the minimal trace length necessary to store all radio pulses and create an empty voltage trace of that length, plus some padding before and after the pulse. Subsequently, the electric field is convolved with the antenna response retrieved from the detector description. The result is then added to the voltage trace, whereby the different trace start times of each electric field trace and the channel's cable delay are taken into account. If the signal stems from an air shower where only one electric field per station is simulated as the signal does not change over such small spacial extents, the differences in signal travel times between channels are also calculated and corrected for. 

\paragraph{Adjusting trace lengths}
If so desired, the lengths of the channel traces can be adjusted using the \emph{channelLengthAdjuster}. If the trace is longer than needed, it is cut to size after the pulse position is determined to ensure it is not accidentally removed. If the trace is too short, it is appended by zeros.

\paragraph{Accounting for amplifiers, filters and cable effects}
Characteristics from the time-dependent detector description (see Sec.~\ref{sec:detector}), are included in different processing modules that convert data to/from ideal voltage traces from/to instrument data. NuRadioReco contains both simplified models, such as ideal filters, and true implementations of the complex behavior of amplifiers and cables. In all these modules, gain and phase-delay is applied to the data. Measurements or simulations of components are interpolated such that they match the sampling rate of the data at that point in the processing step. Also, via the units utilities, the use of different units at different steps of the processing chain is automatically accounted for..

\paragraph{Noise generator}
For a realistic simulation of signals recorded by the antennas it is essential to add noise. This can be done in the module \emph{channelGenericNoiseAdder}. In the module, simple white noise with a normal amplitude distribution for a frequency band specified by the user is calculated in the frequency domain. The user also specifies the required RMS voltage of the noise in the time domain. In most cases, it will be desirable to remove the zero-frequency component, in which case the RMS of the voltage values will be identical to their standard deviation. When calculating the corresponding amplitudes of each frequency bin, it is taken into account that only the specified frequency band contributes to the signal power.

To obtain the desired noise distribution, the phase of each frequency bin is drawn from a uniform random distribution in the range $[0,2\pi)$. The amplitude can be chosen either to be perfectly flat over the specified frequency range or to follow a Rayleigh distribution for each frequency bin. In the time domain, the former yields the specified RMS voltage \emph{exactly}, while in most cases providing a reasonable approximation of the noise background. This mode is hence ideally suited for developing and debugging new code. The Rayleigh distribution is expected for the absolute value of complex numbers, where the real and imaginary components are uncorrelated and each follow a normal distribution with equal variance and zero mean. This provides for a more realistic noise model with small statistical deviations from the specified RMS voltage in the time domain.

\paragraph{Trigger simulations}
The last step in generating real sensitivities is the simulation of a trigger. Per default, several options are included in NuRadioReco. A simplified threshold trigger, an ARIANNA-style dual-thres\-hold trigger, as well as an ARA-style tunnel-diode trigger. The framework can also account for a trigger from the phasing of several antennas, and is capable of calculating coincidence requirements across multiple channels.

\paragraph{Template correlation}
In order to distinguish desired signals from background noise, recorded voltage traces can be correlated with one or more neutrino or cosmic-ray waveform templates using the \emph{channelTemplateCorrelation} module.
The templates are voltage traces generated from simulated radio pulses, which have to be calculated beforehand. The \emph{channelTemplateCorrelation} module re-samples the template to match the sampling rate of the recorded signal and calculates the correlation
\begin{equation}
    \rho (V_{sig}, V_{tmp}, \Delta n) = \frac{\sum_i (V_{sig})_i\cdot (V_{tmp})_{i-\Delta n}}{\sqrt{\sum_i(V_{sig})_i^2}\cdot\sqrt{\sum_i(V_{tmp})_i^2}}
\end{equation}
where $V_{sig}$ and $V_{tmp}$ are the voltage traces of the re\-cor\-ded signal and the template, respectively, which are shifted by $\Delta n$ samples relative to each other. The denominator normalizes the expression to $-1 \leq \rho \leq 1$. The $\Delta n$ yielding the highest correlation is found and both the correlation and the time offset corresponding to $\Delta n$ are saved in the parameter storage. If the channel was compared to multiple templates, the average value over all templates is also calculated, as well as the maximal correlation. 

\paragraph{Directional reconstruction}
Both modules providing a reconstruction of the signal direction, the \emph{correlationDirectionFitter} and the \emph{templateDirectionFitter} use the same principle: Assuming a plane-wave, a signal coming from the direction $\vec{e}_r(\theta,\phi)$ will arrive at an antenna positioned at $\vec{x}$ at the time
\begin{equation}
    t_{exp} = \vec{e}_r(\theta,\phi) \cdot \vec{x} \cdot \frac{n}{c} + t_0 \, ,
\end{equation}
where $c$ is the speed of light in vacuum and $n$ is the index of refraction of the medium surrounding the antennas. 
The \emph{templateDirectionFitter} uses the relative time shift $t_{corr}$ for which the \emph{channelTemplateCorrelation} module found the best correlation to a template and minimizes the $\chi^2$-function
\begin{equation}
    \chi^2 =\sum_i \frac{((t_{exp, i}(\vec{e}_{signal}) - \langle t_{exp}\rangle) - (t_{corr,i} - \langle t_{corr}\rangle))^2}{\sigma_t^2}
\end{equation}
where $<t_{exp}>$ and $<t_{corr}>$ are the averages of $t_{exp}$ and $t_{corr}$, respectively.

The \emph{correlationDirectionFitter} takes two pairs of channels that measure the same polarization and correlates them with each other. It then finds the direction for which correcting for the time difference between channels results in the best correlation between channel pairs. The advantage of this method is that it is independent of the description of the antenna response as only the time differences of parallel channels are considered where the antenna response is the same and thus cancels out as systematic uncertainty.
This method is tailor-made for an ARIANNA-like detector with parallel channels. A more general direction-fitting routine can easily be adapted from the existing modules. Similarly, modules for spherical or hyperbolic arrival times are not in the default repository. 

\paragraph{Converting voltages to electric fields}
After having identified signals and having obtained their arrival direction, a typical task is to reconstruct the electric field from the measured data, essentially inverting the \emph{efieldToVoltageConverter}. In the presence of noise, this is however not a straight-forward inversion and we present two novel algorithms to obtain this from spatially distributed antennas in Sec.~\ref{sec:Reco_algo}. 

\section{Data visualization}
\begin{figure*}[t]
\centering
\includegraphics[width=0.9\textwidth]{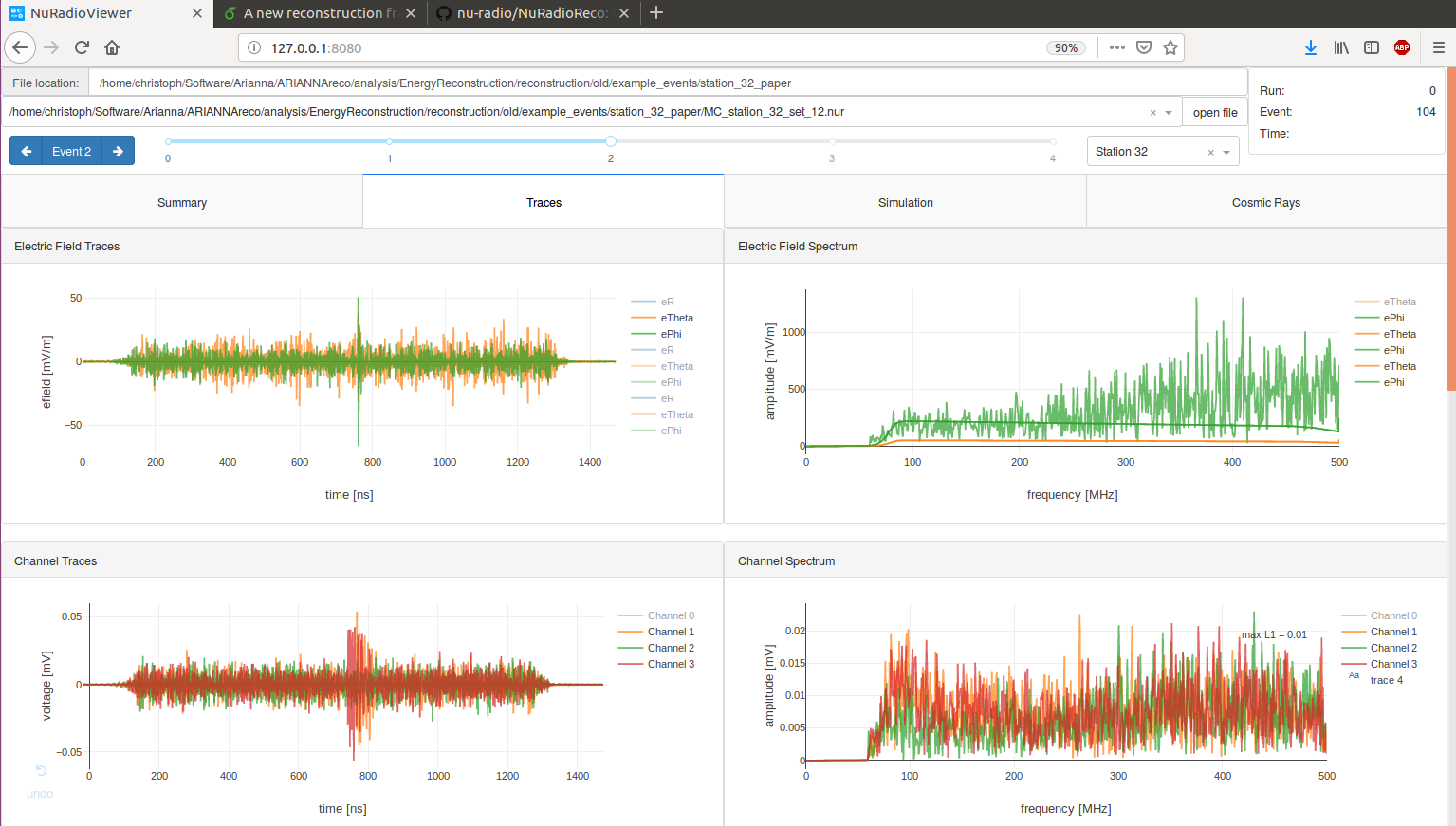}
\caption{Screenshot of the EventBrowser. The EventBrowser can be rendered in any modern web browser. Panels at the top switch between various summary figures (e.g. pulse amplitude as function of time), individual event data and if present simulated and cosmic-ray related data. This example shows the electric field and a fraction of the channel waveforms for illustration. The EventBrowser allows for zooming in and out of very plot individually and figures in different tabs are connected for event selection.}
\label{fig:eventbrowser}
\end{figure*}

NuRadioReco uses state-of-the-art web technologies for data visualization. Using web technologies comes with a number of advantages:
\begin{itemize}
\item The GUI is platform independent and the only requirement is a web browser. Hence, computers and laptops with all operating systems can be used as well as tablets or smartphones. 
\item Data can easily be visualized and made available on the internet, opening up new possibilities for outreach activities and collaborative sharing.
\item HTML templates and CSS spreadsheets provide an efficient way to design responsive user interfaces.
\item The layout and behavior of the user interface can easily be extended by external libraries, such as Bootstrap \cite{bootstrap}. 
\end{itemize}
The NuRadioReco EventBrowser is based on the Dash package \cite{dash}. While Dash itself is written in Python, it creates an HTML/JavaScript template that is rendered by the web browser and can be extended using custom JavaScript or CSS add-ons.

Thanks to this framework, the EventBrowser is responsive and customizable (see Fig.~\ref{fig:eventbrowser}). For example, any graph can be zoomed into or out of.  Additional quantities, such as overview quantities as function of time can be visualized and a mouse click on a point representing a specific event will immediately show the details of the event in the EventBrowser.
The EventBrowser can be used locally by starting a webserver via
\begin{minted}[
%frame=lines,
%framesep=2mm,
%baselinestretch=1.2,
fontsize=\footnotesize,
]{bash}
python NuRadioReco/eventbrowser/index.py /path/files
\end{minted}
with the last argument passed to the python command specifying the location of the data files files to be viewed. The EventBrowser can then be accessed by opening any web browser and going to the address provided in the terminal output, which is http://127.0.0.1:8080/ by default. The EventBrowser lists all .nur files in the specified location in a drop down menu, from which they can be selected for viewing.

\section{New reconstruction algorithms for the electric field}
\label{sec:Reco_algo}

In this section we present two novel reconstruction algorithms for the electric field, developed for the station layout of radio neutrino detectors. The incident electric field is a central quantity as many other properties such as the signal polarization, the energy fluence and the frequency spectrum are directly calculated from it, which are then used to determine the neutrino or cosmic-ray properties. For example, to reconstruct the arrival direction of a neutrino, one naturally needs the signal arrival direction, but also the frequency slope and polarization to determine on which part of the Cherenkov cone the signal was detected. The polarization breaks the degeneracy around the shower axis and the frequency slope determines the angle to the shower axis.
Without information about polarization and frequency slope, the signal could have been detected anywhere on the Cherenkov cone leaving a large uncertainty on the neutrino arrival direction. Also for the energy reconstruction of neutrinos, knowing on which part of the Cherenkov cone the signal was detected is crucial as the Askaryan signal amplitude depends on it. For cosmic-rays, analyses recovering the full electric-field provide the most precise reconstruction of the energy to-date. Hence, an accurate electric-field reconstruction is a crucial parameter for the overall event reconstruction.

Here, we will concentrate on the cosmic-ray case. There is no standard-approach to neutrino reconstruction yet and discussing such strategies in detail would go beyond the scope of this paper. However, one can easily envision how the concept of the forward folding (Sec.~\ref{sec:forwardfolding}) can be applied to neutrino pulses.

Cosmic-ray detectors typically are built with dual-po\-la\-rized antennas, i.e., the radio signal is measured at the same point in space and time in two orthogonal polarizations which allows for a straight-forward reconstruction of the three-dimensional electric field \cite{RadioOffline,AERAPolarization,LofarPolarization2014}.  In contrast, radio neutrino detector stations typically consist of multiple spatially separated antennas of different orientations to maximize the effective volume for neutrino detection and to
minimize the antenna costs. Cosmic-ray reconstruction in neutrino detectors therefore required the development of a new method to reconstruct the incident three-dimensional electric-field pulse. 

\subsection{Standard reconstruction}
\label{sec:standard_reconstruction}

For antenna separations within a detector station of less than \SI{10}{m} (such as the dimensions of a current ARIANNA station) it can safely be assumed that all antennas observe the same pulse generated by an air shower. Using the reconstructed signal arrival direction $(\varphi_0,\vartheta_0)$, we correct for the time delays and combine the measurements of all antennas into an over-determined system of equations that is solved for the electric field using a chi-square minimization per frequency bin.

Mathematically, this is expressed in the frequency domain as 
\begin{equation}
    \begin{pmatrix} \mathcal{V}_1(f) \\ \mathcal{V}_2(f) \\ ...\\ \mathcal{V}_n(f)\end{pmatrix} = 
    \begin{pmatrix} \mathcal{H}_1^\theta (f)& \mathcal{H}_1^\phi (f)\\ \mathcal{H}_2^\theta (f) & \mathcal{H}_2^\phi (f)\\ ... \\ \mathcal{H}_n^\theta (f)& \mathcal{H}_n^\phi (f)\end{pmatrix} 
    \begin{pmatrix} \mathcal{E}^\theta(f) \\ \mathcal{E}^\phi(f)\end{pmatrix} \, ,
    \label{eq:H_full}
\end{equation}
where $\mathcal{V}_i$ is the Fourier transform of the measured voltage trace of antenna $i$, $\mathcal{H}_i^{\theta, \phi}$ represents the response of antenna $i$ to the $\phi$ and $\theta$ polarization of the electric field $\mathcal{E}^{\theta, \phi}$ from the direction $(\varphi_0,\vartheta_0)$. This system of equations is then solved for $\mathcal{E}^{\theta, \phi}$. Due to the typical noise contribution on measured waveforms, there is no perfect solution. Hence, we determine the electric field values $\mathcal{E}^{\theta}(f_i)$, $\mathcal{E}^{\phi}(f_i)$ for each frequency bin $f_i$ that minimize the sum of the squared differences of $\mathcal{V}_i(f_i)$.

We refer to this technique as the \emph{standard technique} as it is an extension of the method used in dedicated cosmic-ray detectors from two orthogonal channels to many. In comparison to the former, it has the advantages that the signal-to-noise ratio is reduced by adding more antennas to the reconstruction, and, more importantly, it allows for more flexible station designs. For example, instead of building a complicated 3D antenna that measures all three electric-field polarizations at the same point in space, one could simply place a dedicated vertical antenna at a few meters distance from a dual-polarized antenna and combine the signals in software rather than hardware.

The standard method has proven to work reliably as long as at least two orthogonal antennas have a good signal-to-noise ratio in all frequency bins. However, there are two shortcomings. The general assumption of this deconvolution method is that all measured voltages originate only from the incident electric field, while in reality it is a sum of the electric-field signal and recorded noise. In a scenario, where an electric-field pulse has no high-frequency content and the upper half of the bandwidth is therefore dominated by noise, this reconstruction method produces incorrect results at high frequencies.  

A second limitation of this method occurs if the signal is only measured in two of three orthogonal polarization components, which is the case for most radio cosmic-ray detectors that only measure the two horizontal (east-west and north-south) components. Although this is in principle sufficient information to determine the full electric field, using the signal arrival direction, the algorithm leads to incorrect results for horizontal air showers \cite{GlaserPhD2017}: Here, the $\vec{e}_\theta$ component of the electric field has a strong vertical and only a small horizontal component. If the antennas are only sensitive to the small horizontal component, the signal is often below the noise level. In the reconstruction it is assumed that the measured noise level is identical to the horizontal component of a much larger $\vec{e}_\theta$ component, leading to a vast overestimation of this polarization component. Thus, current analyses only use the horizontal components of the reconstructed electric field \cite{AERAHorizontal2018}, which does not allow for a proper determination of the polarization or the total signal strength. 

\subsection{Forward folding technique}
\label{sec:forwardfolding}

\begin{figure*}
\centering
\includegraphics[width=1\textwidth]{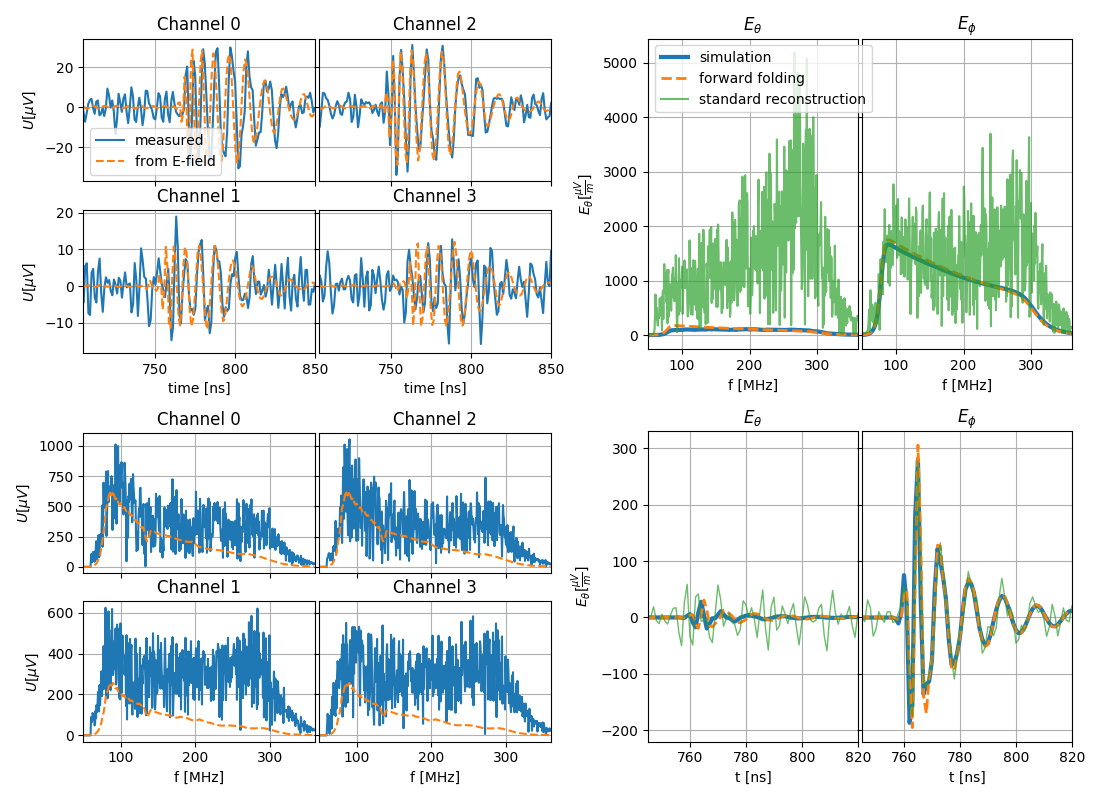}
\caption{
Example of the electric-field reconstruction using the standard and the forward folding technique. Left panels: Voltage traces of four spatially displaces antennas. Shown are both the time- (top) and frequency domain (bottom). The solid blue curve represents the measured voltages whereas the dashed orange curve shows the analytic solution of the forward folding technique. Channels 0,2 and 1,3 are parallel, the measured signal only differs in noise contribution. Upper right panels: Reconstructed amplitude spectrum using the forward folding (dashed orange) and standard (solid green) technique in comparison with the simulated truth (solid blue). Lower right panels: Reconstructed electric field trace using both techniques in comparison to the simulated true values (same colors as above).}
\label{fig:deconvolution}
\end{figure*}

We developed the \emph{forward folding technique} to reconstruct the electric field from multiple channel measurements and address the shortcomings of the \emph{standard method}. It improves the reconstruction for small signal-to-noise ratios, can be used for horizontal showers and prevents spurious results for cases in which the bandwidth of the signal is smaller than the detector bandwidth. Early versions of this technique were already presented in \cite{GlaserPhD2017,GlaserARENACR2018}. 

Instead of recovering numerically the incident electric field, frequency bin by frequency bin, we fit an analytic model of the electric-field pulse directly to the measured voltages in the time domain. Here, we concentrate on the application of this technique for cosmic-ray signals. An extension to neutrino Askaryan pulses should be straight forward and will be subject to forthcoming studies. 

In a typical experimental bandwidth, a cosmic-ray radio pulse can be described sufficiently well with just four parameters in the frequency domain: the signal amplitude of both polarization components $A_{\theta, \phi}$, the frequency slope $m_f$ and a phase offset $\Delta$\footnote{As all signals in the time domain are real-valued, only the positive frequencies are considered. The amplitudes of the negative frequencies are just the complex conjugates of the amplitudes of the positive frequencies. Hence, the phase offset is $-\Delta$ for negative frequencies.} (see Eq.~\eqref{eq:pulse}). It should be noted that this assumes that the pulse is fully linearly polarized. Should one want to study the small component of circular polarization \cite{2016PhRvD..94j3010S}, an adaptation is needed. For additional discussion refer to \cite{GlaserPhD2017}.

We forward fold the such parameterized pulse with the antenna responses of the different channels by multiplying $\mathcal{E}^{\theta, \phi}(A_{\theta, \phi}, m_f, \Delta)$ with the antenna response according to Eq.~\eqref{eq:H_full}. The resulting voltage traces are compared to the measurement. The optimal parameters of the electric-field pulse $(A_{\theta, \phi}, m_f, \Delta)$ are determined in a chi-square minimization in the time-domain of all channels simultaneously.

By applying the antenna response in forward direction to a noiseless waveform, the effects of noise are minimized and an artificial overestimation of the signal is reduced. The fundamental principle of comparing prediction to measurement in the instrumental voltages and not in the physical quantity, the electric field, has already been used in the LOFAR analysis \cite{LOFARPRD}. Their approach of using many dedicated CoREAS simulations per event, is however significantly more computationally expensive and only used to reconstruct high quality detections. 

One example of an electric-field reconstruction presented in Fig.~\ref{fig:deconvolution} illustrates the two main advantages of the forward folding technique: A small amplitude in one of the polarization components is correctly identified (the $\theta$- component in this example), and high frequency components, where the signal amplitude is smaller and the antenna has a reduced sensitivity, are not overestimated. The recovered electric field is less biased by noise. 

\subsection{Implementation details}

The electric field pulse of a cosmic ray is described in the frequency domain as
\begin{equation}
     \begin{pmatrix} \mathcal{E}_\theta \\ \mathcal{E}_\phi \end{pmatrix} = \begin{pmatrix} A_\theta \\ A_\phi \end{pmatrix} 10^{f \cdot  m_f} \, \exp(\Delta\, j) \, 
\label{eq:pulse}
\end{equation}
where $f$ is the frequency and $j$ stands for the imaginary unit. The four parameters that describe the electric-field pulse are: the amplitudes of the $\theta$ and $\phi$ components $A_{\theta, \phi}$, the frequency slope $m_f$ and the phase offset $\Delta$. We do not consider a (linear) phase slope because it corresponds to a shift of the pulse position in the time domain and we assume that time differences have been removed by correcting for the reconstructed arrival direction. 

We found that the following incremental fitting procedure leads to stable results: First only the frequency slope $m_f$ is determined. We use the sum of the maximum cross correlation of all participating channels as objective function. 
\begin{equation}
    \sum_i -\max\left(\rho\left[V_i, \mathrm{IFFT}\left(\begin{pmatrix} \mathcal{H}^\theta_i & \mathcal{H}^\phi_i \end{pmatrix} \begin{pmatrix} \mathcal{E}_\theta \\ \mathcal{E}_\phi \end{pmatrix}\right)\right]\right) \, ,
\end{equation}
where $V_i$ is the measured voltage trace of channel $i$, $\mathcal{H}^{\theta, \phi}_i$ is the antenna vector effective length, $\mathrm{IFFT}$ represents an inverse fast Fourier transform and $\rho$ is the normalized Pearson correlation defined as 
\begin{equation}
    \rho(x, y)_k = \frac{\sum_n x_{n+k} y_n}{\sqrt{\sum_k x_k^2} \sqrt{\sum_k y_k^2}}
\end{equation}
for each time bin $k$. This objective function conveniently removes the dependence on the amplitude and determines the time shift of the analytic pulse with respect to the measured voltages which is given by the value $k$ that maximizes $\rho(x, y)$. In this first optimization step, we set $\Delta = 0$, $A_\phi = 1$ and $A_\theta$ = 0, because cosmic-ray signals are mostly polarized in $\vec{e}_\phi$-direction due to the orientation of the geomagnetic field in polar regions.

In the next iteration, the amplitude is determined by minimizing the following objective function:
\begin{equation}
    \chi^2 = \sum_i \left(\sum_k \frac{\left|V_{i,k} - \mathrm{IFFT}\left(\begin{pmatrix} \mathcal{H}^\theta_i & \mathcal{H}^\phi_i \end{pmatrix} \begin{pmatrix} \mathcal{E}_\theta \\ \mathcal{E}_\phi\end{pmatrix} \right)_k\right|}{V_\mathrm{RMS}}\right)^2 \, ,
    \label{eq:obj2}
\end{equation}
where the index $i$ runs over the channels, $k$ runs over the time bins of each channel and $V_\mathrm{RMS}$ is the RMS of the measured voltage traces. In a first step, only $A_\phi$ is determined, $A_\theta$ and $\Delta$ are set to zero, and $m_f$ is fixed to the previous fit result. In the second step, both $A_\phi$ and $A_\theta$ are optimized simultaneously. 

In the final step, the amplitudes and the slope parameter are optimized simultaneously using the objective function of Eq.~\eqref{eq:obj2} but using the Hilbert envelope of the voltage traces instead of the voltage traces directly. 

\begin{figure*}
\centering
\includegraphics[width=0.8\textwidth]{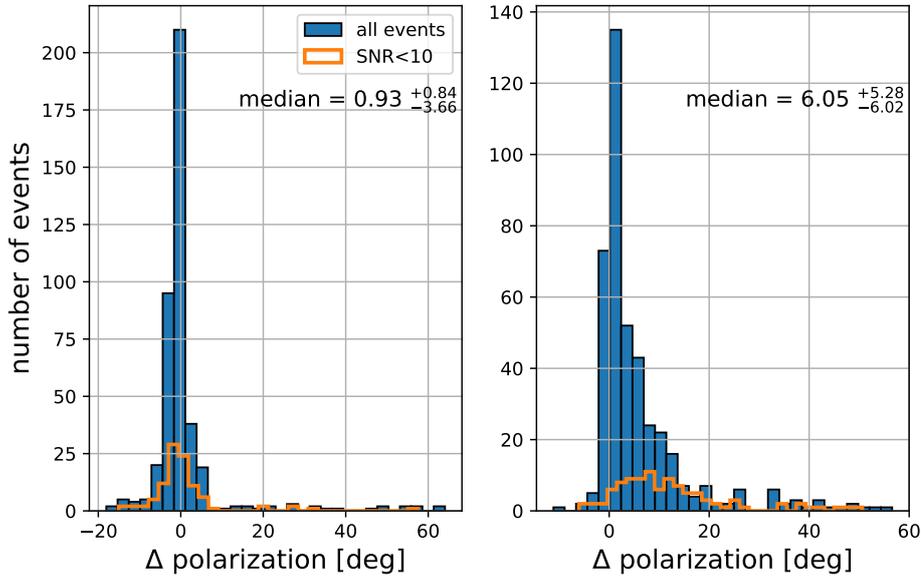}
\caption{Resolution of the reconstructed polarization forward folding (left) and standard method (right). Shown is the angular difference between true polarization and reconstructed polarization. The orange line indicates events with signal-to-noise ratios smaller than 10. For the forward folding method, the distribution's median is $(0.93^{+0.84}_{-3.66})^{\circ}$ and for the standard method it is $(6.05^{+5.28}_{-6.02})^{\circ}$, with the uncertainties specifying the 68\% quantiles.}
\label{fig:polarization_reconstruction}
\end{figure*}

\subsection{Performance of electric field reconstruction}

\begin{figure*}
    \centering
    \includegraphics[width=0.7\textwidth]{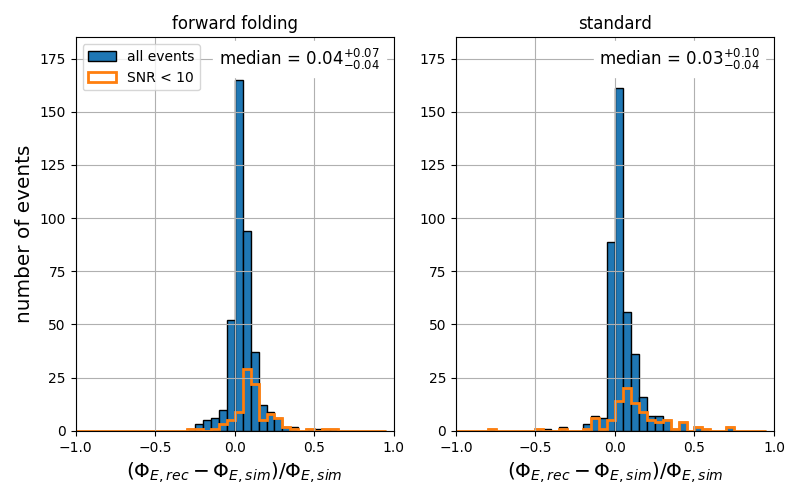}
    \includegraphics[width=0.7\textwidth]{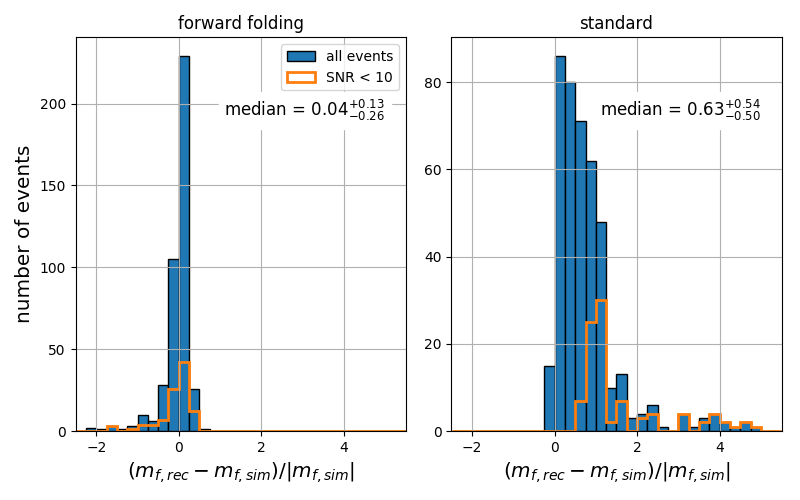}
    \caption{Top row: Resolution of the reconstructed energy fluence using the forward folding (left) and standard methods (right). The orange line indicates events with signal-to-noise ratios smaller than 10. For the forward folding method, the distribution's median is $0.04^{+0.07}_{-0.04}$ and for the standard method it is $0.03^{+0.10}_{-0.04}$, with the uncertainties specifying the 68\% quantiles. 
    Bottom row: Relative uncertainty on the reconstructed frequency slope $m_f$ using the forward folding (left) and standard (right) method. The orange line indicates events with signal-to-noise ratios smaller than 10. For the forward folding method the distribution has a median of $0.04^{+0.13}_{-0.26}$ and for the standard method a median of $0.63^{+0.54}_{-0.50}$, with the uncertainties specifying the 68\% quantiles. }
    \label{fig:frequency_slope_hist}
\end{figure*}

We evaluate the performance of the electric-field reconstruction using the standard and the forward folding technique in a Monte Carlo study with CoREAS air shower pulses. We calculate the polarization, the energy fluence and the frequency slope from the reconstructed electric field and compare it to the Monte Carlo truth. 
In case of the forward folding technique, the polarization is given by $\arctan(A_\theta/A_\phi)$, defining it as the angle between $\vec{e}_\phi$ and the electric field vector. The frequency slope is given by the parameter $m_f$ (see Eq.~\ref{eq:pulse}).

We perform the Monte Carlo study for a detector layout consisting of four upward facing LPDA antennas where one pair is oriented along the North-South direction and the other pair is oriented along the East-West direction. All antennas are placed at a distance of \SI{4}{m} to the center of the station. This corresponds to the station layout of a dedicated cosmic-ray station of the ARIANNA detector. We note that the method works with any kind of antenna but we focus on LPDAs here because the cosmic-ray signal is mostly horizontally polarized and vertically aligned dipoles typically do not add signal.

We present the resolution for the reconstructed polarization for a set of 100 simulated air showers in Fig.~\ref{fig:polarization_reconstruction}. The data set is generated to resemble a realistic cosmic-ray distribution. We also use the module \emph{readCoREAS}, which randomly picks shower core positions. For each simulated shower 150 randomly chosen shower core positions are used. The energy distribution for the events follows a power-law with spectral index --2 and the arrival directions are isotropically distributed. We apply a full detector simulation and event reconstruction as detailed in the example of Sec.~\ref{sec:example_rec}, including the simulation of noise with an RMS amplitude of \SI{20}{mV}. 

We consider only events that exceed a trigger threshold of at least \SI{100}{mV} in two of the channels. An additional cut is made on the signal-to-noise ratio\footnote{SNR is defined as half of the peak-to-peak amplitude divided by the noise RMS} for all channels, which needs to be higher than 4. This cut is needed because the current arrival direction reconstruction fails when there is no detectable signal in one of the two channel-pairs. It selects 423 pulses out of 603 pulses obtained by the event selection. The event set selected by this cut still contains 95\% of the events for which the directions are considered to be well reconstructed, meaning a direction reconstruction within 5 degrees of the simulated direction. For the electric-field reconstruction, a frequency range of 80 - 500 MHz is used.

The obtained resolution of the polarization, energy fluence and frequency slope are presented in Figs.~\ref{fig:polarization_reconstruction} and \ref{fig:frequency_slope_hist}.
For the polarization reconstruction using the standard method, we obtain a median resolution of $(6.05^{+5.28}_{-6.02})^{\circ}$. 
For the forward folding method, we obtain a median of $(0.93^{ +0.84}_{-3.66})^{\circ}$.  The uncertainties represent the 68\% quantiles. The better performance for the forward folding method is especially visible for the low signal-to-noise ratio events, which are indicated by the orange histogram in the figure (SNR \textless 10). Overall, the method indeed addresses the shortcomings observed earlier. 
 
In the reconstruction of the energy fluence of the radio signal, standard and forward folding methods perform equally well, as shown in Fig.~\ref{fig:frequency_slope_hist}, which shows the relative accuracy of the methods. With the standard methods, we obtain a median relative uncertainty of $0.03^{+0.10}_{-0.04
}$ and with the forward folding a median of $0.04^{+0.07}_{-0.04}$. The reason for a similar performance is that the average noise contribution is subtracted from the energy fluence that is calculated as integrated quantity around the signal pulse, making the energy fluence a robust estimator in the presence of noise \cite{AERAPRD}. 

The quality of the reconstruction of the frequency slope parameter $m_f$ is shown in Fig.~\ref{fig:frequency_slope_hist}. For the simulated electric field and the one reconstructed using the standard method, $m_f$ was determined by a linear fit to $\log_{10}(|\vec{E}(f)|)$, while for the forward folding method the parameter was a result of the fitting process.

The forward folding method shows a solid performance, resulting in a relative uncertainty with a median of $0.04^{+0.13}_{-0.26}$, while the standard method performs much worse. Especially for signals with a low SNR, $m_f$ is overestimated. As described in Sec. \ref{sec:standard_reconstruction}, noise can cause the electric field strength at higher frequencies to be overestimated, leading to a value of $m_f>0$, while the actual $m_f$ is almost always negative in this frequency range. For this reason, the resulting distribution has a median of $0.63^{+0.54}_{-0.50}$, where the uncertainties represent the 68\% quantiles.

To summarize, in all three quantities tested, polarization, energy fluence, and frequency slope, the forward folding performs on-par or better than the standard method. Especially for reconstructions sensitive to the polarization and the frequency slope, we recommending using the forward folding method to recover the electric field and to overcome biases and reduce the contribution of noise. 

\section{Conclusions}
We have presented a new Python-based re\-con\-struc\-tion framework for particle radio detectors. The framework has been designed to analyze data from current radio neutrino detectors, such as ARIANNA, and to prepare the reconstruction for a planned large radio array. Due to the design-goal of high flexibility it is also usable for cosmic-ray radio detectors. The framework provides both a native data structure and a time-dependent detector description, which is designed to account for large and complex detectors. Data visualization relies on web-based tools, allowing for the easy separation of a server-based data analysis and remote inspection. 

In its current version, the framework provides the algorithms for all steps necessary to reconstruct the full electric field for incoming cosmic-ray signals and basic event identification for neutrino detectors. A well-documented example is provided in the code. Due to the strict modularity it is straight-forward to design additional modules to complete the reconstruction for neutrinos. Also, previously unpublished reconstruction algorithms are made available to the community in the code and have been described in this article. 

As this framework builds on experience gained with all currently used software for the radio detection of cosmic rays and neutrinos, we were able to anticipate a number of complexities and avoid them in software design. We expect the library of standard algorithms available in NuRadioReco to grow in time along with their development in the neutrino community. 

\section{Acknowledgements}
We would like to thank the members of the InIceMC working group, consisting of members from the ARA and ARIANNA collaborations, for helpful discussions regarding simulations of the radio signal of neutrinos that helped shape our reconstruction framework. In particular, we would like to thank Simon Archambault and Keiichi Mase from Chiba University for discussions concerning the handling of antenna responses in different simulation tools.  

We acknowledge funding from the German research foundation (DFG) under grants GL 914/1-1 and NE 2031/2-1, and the U.S. National Science Foundation-Physics Division (grant NSF-1607719).

\bibliographystyle{spphys}
\bibliography{BIB}

\appendix
\section{Extracting the realized effective length from antenna simulations}
\label{sec:Antenna_effective}
For completeness we provide the equations used to calculate antenna quanitities for the detector simulation. This hopefully allows the reader to compare different antenna simulations programs such as WIPL-D and XFDTD. 
\subsection{The vector effective length}

The vector effective length $\vec{\mathcal{H}}$ relates the incident electric field $\vec{\mathcal{E}}$ to the open circuit voltage at the antenna terminals $\mathcal{V}_\mathrm{OC}$ (in the Fourier domain) as
\begin{equation}
\mathcal{V}_\mathrm{OC} = \vec{\mathcal{H}}\cdot \vec{\mathcal{E}} = (\mathcal{H}_\theta, \mathcal{H}_\phi) \cdot (\mathcal{E}_\theta, \mathcal{E}_\phi)^T
\end{equation} 
where $\vec{\mathcal{H}}$ and $\vec{\mathcal{E}}$ are vectors in spherical coordinates, having $\mathcal{H}_r = \mathcal{E}_r = 0$, thus only having signal in the $\theta$ and $\phi$ polarization.

In a measurement setup the antenna will be read out at a load impedance. For a simple measurement setup we get
\begin{equation}
\mathcal{V}_L = \frac{Z_L}{Z_A + Z_L} \mathcal{V}_\mathrm{OC} \,,
\end{equation}
where $Z_L$ is the load impedance which is typically \SI{50}{\ohm} and $Z_L$ is the antenna impedance.
The realized VEL is then given by
\begin{equation}
\vec{\mathcal{H}}_{rl} = \frac{Z_L}{Z_A + Z_L} \vec{\mathcal{H}} \, .
\label{eq:Hr2}
\end{equation}
and we can relate the incident electric field to the measured voltage by
\begin{equation}
\mathcal{V}_\mathrm{L} = \vec{\mathcal{H}}_{rl}\, \vec{\mathcal{E}} \, .
\end{equation} 

\subsection{Simulation of the effective height}
Most antenna simulation software computes the far field electric field generated by an antenna, i.e., they simulate an emitting antenna and due to reciprocity the receiving antenna case can also be calculated from such a simulation. The vector electric field $\vec{\mathcal{E}}(\omega)$ emitted by an antenna is related to the vector effective length (VEL) as (see \cite{AntennaPaper} Eq.~5.1 or \cite{Kravchenko2007} Eq.~6)
\begin{equation}
\vec{\mathcal{E}}(\omega) = -i Z_0 \frac{1}{2 \lambda R} \mathcal{I}_0 \vec{\mathcal{H}} \exp(-i \omega R/c) \, .
\label{eq:1}
\end{equation}
where $Z_0$ is the free space impedance, $\lambda$ is the wavelength, $R$ is the distance to the antenna and $\mathcal{I}_0$ is the current at the feedpoint of the antenna.

This equation can be simplified by normalizing the electric field to a unit distance at \SI{1}{m}, which removes the distance dependence from the equation. The normalized electric field (also called complex voltage) is given by
\begin{equation}
\vec{\mathcal{E}}'(\omega)=\vec{\mathcal{E}}(\omega) \, R \, \exp(i \omega R/c) \, .
\label{eq:I}
\end{equation} 
Then, solving for the effective length gives:
\begin{equation}
\vec{\mathcal{H}} = \frac{2 \lambda \mathcal{I}_0}{-i Z_0} \vec{\mathcal{E}}'(\omega) \, .
\label{eq:H3}
\end{equation}

In the case of WIPL-D, not a current $\mathcal{I}_0$ is simulated at the antenna feedpoint but a perfect voltage generator of $V_{OC}$ = 1 Volt. Then,
\begin{equation}
    \mathcal{I}_0 = V_{OC} / Z_A
\end{equation}
and Eq.~(\ref{eq:H3}) becomes
\begin{equation}
\vec{\mathcal{H}} = \frac{2 \lambda Z(\omega)}{-i Z_0 V_{OC}} \vec{\mathcal{E}}'(\omega) \, .
\label{eq:H}
\end{equation}
Then, using Eq.~(\ref{eq:Hr2}) and exploiting the identity $Z_A = \frac{1 + S11}{1-S11} Z_L$ we find the following simplified formula for the realized effective length
\begin{equation}
    \vec{\mathcal{H}_{rl}} = \frac{\lambda (1 + S11) Z_L}{-i Z_0 V_{OC}} \vec{\mathcal{E}}'(\omega) \, .
    \label{eq:Hr}
\end{equation}

\subsection{Relation between realized vector effective length and realized gain}
Gain and realized gain are related via (from Eq. 6.8 of \cite{IEEE_ht})
\begin{equation}
G_{rl}(\omega) = G(\omega)\left[1-|S11(\omega)|^2\right]\, .
\label{eq:GGr}
\end{equation}
The vector effective length (not $\vec{\mathcal{H}_{rl}}$ but $\vec{\mathcal{H}}$) is related to the gain via (from Eq.~(A.8) from \cite{AntennaPaper})
\begin{equation}
|\vec{\mathcal{H}}|^2 = \frac{c_0^2}{f^2\, n}  \frac{\Re(Z_A)}{\pi Z_0} G(\omega) \, .
\label{eq:HG}
\end{equation}
Finally, the realized vector effective length is related to the realized gain via
\begin{equation}
|\vec{\mathcal{H}}_{rl}|^2 = \frac{c_0^2}{f^2\, 4 \, n}  \frac{\SI{50}{\Omega}}{\pi Z_0} G_{rl}(\omega)\, .
\label{eq:HrGr}
\end{equation}

\subsection{Impact of embedding an antenna in a medium}
We characterize a medium (e.g. ice) by its relative permittivity ($\epsilon_r$), its relative permeability ($\mu_r$) and its conductivity ($\sigma$). In the WIPL-D simulation we set the conductivity to $\sigma$ = \SI{1e-6}{S/m} and assume that $\mu_r = 1$. Then, 
\begin{equation}
    \epsilon_r = n^2 \, ,
\end{equation}
where $n$ is the index of refraction. Also the wave im\-pe\-dan\-ce changes accordingly
\begin{equation}
    Z = \sqrt{\frac{\mu}{\epsilon}} = \sqrt{\frac{\mu_0 \mu_r}{\epsilon_0 \epsilon_r}} \stackrel{\mu_r = 1}{=} Z_0 \sqrt{\frac{1}{\epsilon_r}} = \frac{Z_0}{n} \,
    \label{eq:Z}
\end{equation}
where $Z_0$ is the free space impedance. Thus, the formula to calculate the vector effective height from the simulation output (Eq.~\ref{eq:H}) does not change: The wavelength changes with $n$ by $\lambda = \lambda_0 / n$ but at the same time Eq.~\ref{eq:Z} adds another factor of $n$ that cancels the first. 

\begin{figure}
    \centering
    \includegraphics[width=0.47\textwidth]{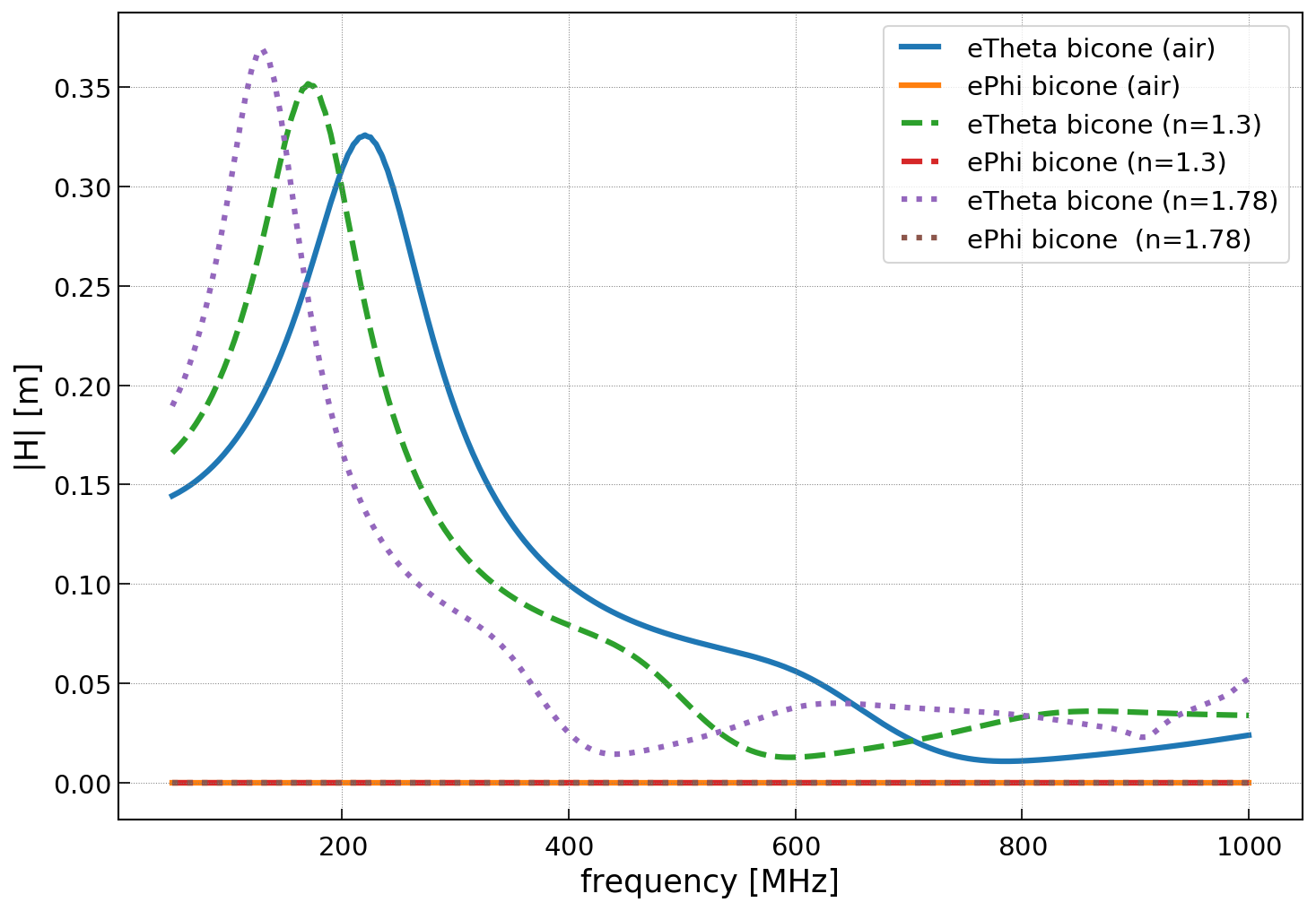}
    \caption{The (open circuit) vector effective length of a WIPL-D simulation of the 'birdcage' bicone antenna \cite{ARA} in different media (air and ice with different refractive indices (n)). }
    \label{fig:bicone_media}
\end{figure}

In Fig.~\ref{fig:bicone_media}, the effective height of a bicone antenna in different media is presented. The complete curve shifts to lower frequencies according to $f = f_{air} / n$ whereas the magnitude of the effective height remains essentially the same. This is in accordance with our intuition that the antenna should behave essentially the same and only the resonance frequency should change according to the change in wavelength. 

\subsection{Useful identities}
The S11 parameter (measured in a \SI{50}{\ohm} system) is related to the antenna impedance by
\begin{equation}
    Z = \frac{1 + S11}{1-S11}  \SI{50}{\Omega}  \iff S11 = \frac{Z -  \SI{50}{\Omega}}{Z +  \SI{50}{\Omega}}
    \label{eq:S11Z}
\end{equation}
The voltage standing wave ratio (VSWR) is related to S11 by
\begin{equation}
    \text{VSWR} = \frac{1 + |S11|}{1-|S11|}
\end{equation}

\end{document}